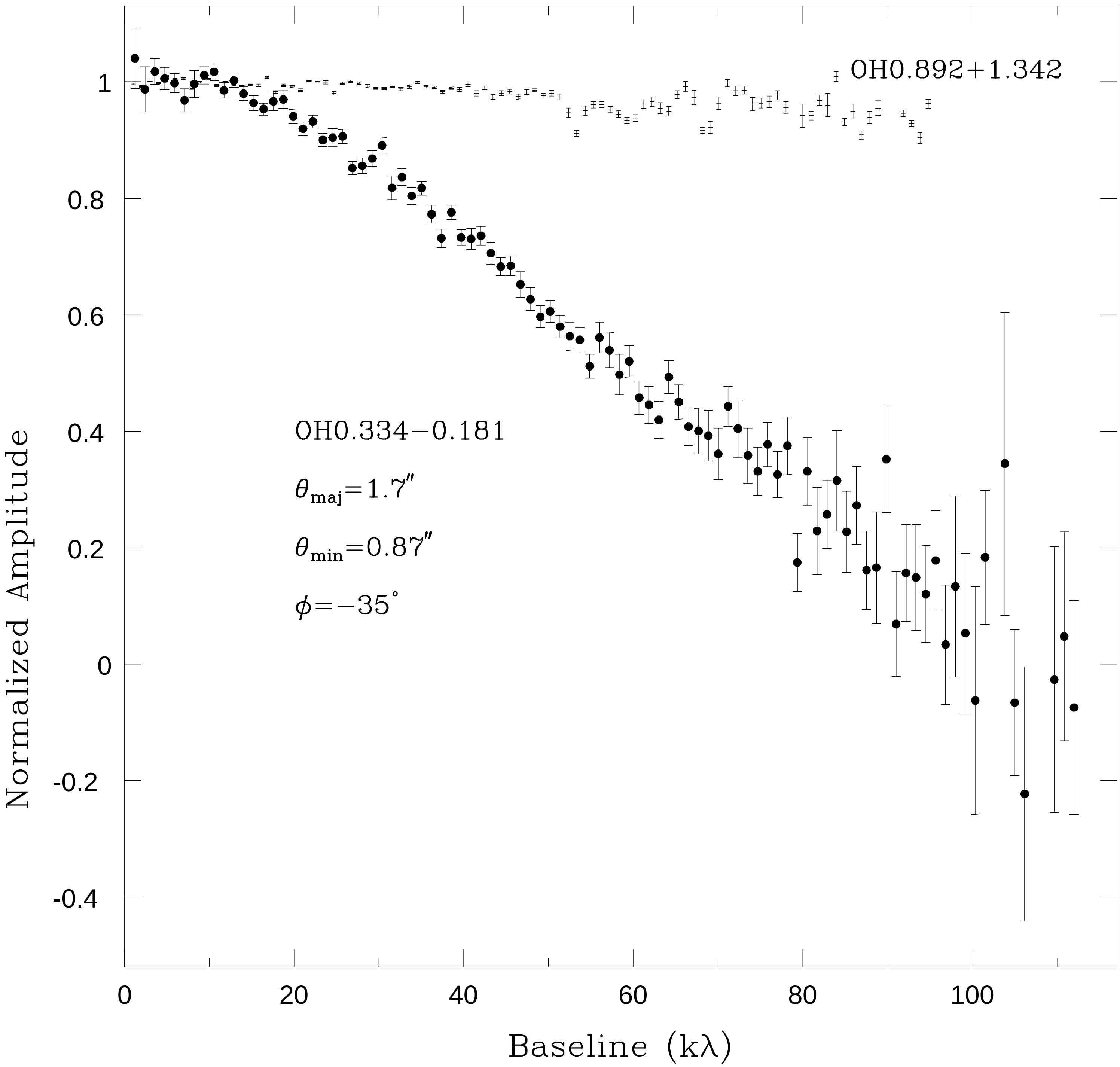

OH0.892+1.342
OH0.334−0.181
$\theta_{maj}$=1.7″
$\theta_{min}$=0.87″
$\phi$=−35°
Normalized Amplitude
Baseline (kλ)
1
0.8
0.6
0.4
0.2
0
-0.2
-0.4
0
20
40
60
80
100

Scattering Model Fit to Visibility Data

| Name<br>(1) | $S_\circ$<br>(2) | $\theta_{maj} \times \theta_{min}$ ($\phi$)<br>(3) | $\epsilon_{\theta_{maj}} \times \epsilon_{\theta_{min}}$ ($\epsilon_\phi$)<br>(4) | $\eta$<br>(5) | $\Delta\Theta$<br>(6) |
|---|---|---|---|---|---|
| OH 0.892+1.342 ...... | 56 Jy | $0''.017$ | $0''.005$ | 1.0 | $100'.7$ |
| OH 0.319−0.040 ...... | 3.8 Jy | $1''.2$ | $0''.05$ | 1.0 | $22'.5$ |
| OH 0.334−0.181 ...... | 4.3 Jy | $1''.7 \times 0''.87$ ($-35°$) | $0''.04 \times 0''.02$ ($2°$) | 2.0 | $24'.7$ |
| OH 359.517+0.001 ...... | 1.3 Jy | $1''.2 \times 0''.56$ ($-11°$) | $0''.05 \times 0''.12$ ($1°$) | 2.1 | $25'.8$ |
| OH 359.762+0.120 ...... | 5.9 Jy | $1''.2 \times 0''.36$ ($+10°$) | $0''.03 \times 0''.11$ ($2°$) | 3.3 | $14'.8$ |
| OH 359.581−0.240 ...... | 4.7 Jy | $1''.3 \times 0''.32$ ($-24°$) | $0''.02 \times 0''.09$ ($1°$) | 4.1 | $24'.7$ |
| OH 359.880−0.087 ...... | 7.8 Jy | $1''.2 \times 0''.94$ ($+81°$) | $0''.03 \times 0''.03$ ($4°$) | 1.3 | $4'.6$ |
| OH 359.986−0.061 ...... | 1.0 Jy | $1''.6 \times 0''.56$ ($-52°$) | $0''.2 \times 0''.45$ ($10°$) | 2.9 | $2'.6$ |

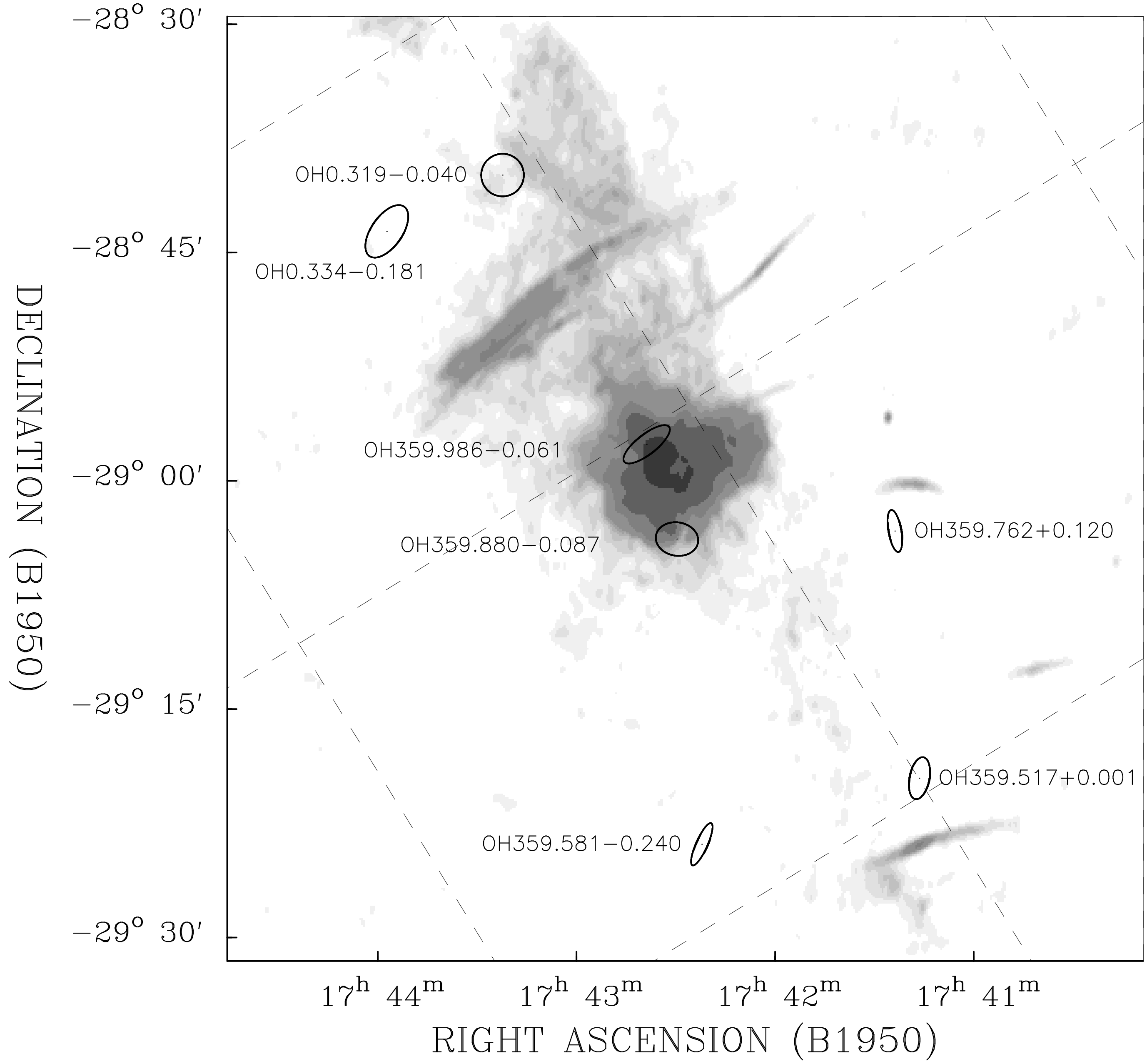

OH0.319−0.040
OH0.334−0.181
OH359.986−0.061
OH359.880−0.087
OH359.762+0.120
OH359.517+0.001
OH359.581−0.240
−28° 30′
−28° 45′
−29° 00′
−29° 15′
−29° 30′
DECLINATION (B1950)
17h 44m
17h 43m
17h 42m
17h 41m
RIGHT ASCENSION (B1950)

# Anisotropic Scattering of OH/IR Stars Towards the Galactic Center

Dale A. Frail *NRAO/AOC, Socorro, New Mexico, U.S.A.*

Philip J. Diamond *NRAO/AOC, Socorro, New Mexico, U.S.A.*

James M. Cordes *Astronomy department and N.A.I.C., Cornell University, Ithaca, New York, U.S.A.*

Huib Jan Van Langevelde *Sterrewacht Leiden, Leiden, The Netherlands*

**Summary.** We have made angular broadening measurements in the satellite line of the OH molecule at 1612 MHz of 8 OH/IR stars located near the Galactic Center with the Very Large Array. Early observations with the Very Long Baseline Array had revealed a region of pronounced scattering approximately centered on SgrA*. These new observations show that the compact line emission from these stars is scattered by an amount that is two to three times larger than had been previously seen. Additionally, we measure non-circular scattering disks for these objects, evidence that the electron density variations are distributed anisotropically. The present data suggest that the enhanced turbulence is located near the Galactic Center. However, we cannot completely rule out that the scattering is caused by an unrelated region along the line of sight.

## 1. Introduction

Early in the study of the Galactic center (GC) it became apparent that our view of this region was "blurred" by plasma turbulence (Davies, Walsh & Booth 1976). Attempts to image the radio source SgrA* showed that its observed angular diameter scales quadratically with wavelength ($\theta_{obs} \propto \lambda^2$) from 1 GHz to at least 43 GHz (see Backer et al. 1993 and references therein). This result is expected if the size of SgrA* is set by interstellar scattering (ISS) (Backer 1988). Important support for this idea came from Van Langevelde & Diamond (1991), who showed that the 1612 MHz OH masers in the circumstellar shells of four OH/IR stars near SgrA* were broadened by a similar amount. Van Langevelde et al. (1992, hereafter Paper I) followed up this work with a more detailed study on the distribution of ISS towards the Galactic center. They found a region of pronounced scattering ($\bar{\theta}_{obs} \simeq 525$ mas at 1.6 GHz) approximately centered on SgrA*, but were unable to determine whether the scattering originated at the Galactic center or due to a random clump of enhanced turbulence along the line of sight. For 7 of the 20 OH/IR stars studied in Paper I there were no fringes detected on any interferometer baselines of the Very Long Baseline Array (VLBA) [1] . It was suggested that

---

**Note 1:** The Very Long Baseline Array (VLBA) and the Very Large Array (VLA) are operated by the National Radio Astronomy Observatory under cooperative agreement with the National Science Foundation.

---

this was because these lines of sight were scattered further still and thus the OH/IR stars were resolved even by the shortest VLBA baselines. In order to confirm this prediction we obtained VLA[1] observations of several of these OH/IR stars. The new measurements that we report here supersede an earlier, more limited dataset published by Diamond et al. (1993).

## 2. Observations

The observations were made at the VLA[1] on 1993 January 30 and 31 in the hybrid BnA array. A bandwidth of 781 kHz and 64 frequency channels were recorded in each hand of circular polarization for a resolution of 12.2 kHz per channel. Four OH/IR stars from Paper I were observed on each day, one of which had a previously determined VLBA angular diameter measurement while the remainder had only lower limits.

The data were calibrated within the AIPS package following standard practice. The complex visibility data $\mathrm{V}(u, v)$ for the peak spectral line channel were fitted by a 6-parameter elliptical gaussian model of the form,

$$\mathrm{V}(u, v) = \mathrm{S}_{\circ}\ exp\left[-\left(\frac{u_{\phi}^{2}}{A^{2}} + \frac{v_{\phi}^{2}}{B^{2}}\right)\right]\ exp\left[\frac{2\pi}{\lambda} i\left(\mathrm{X}_{u} + \mathrm{Y}_{v}\right)\right], \tag{1}$$

where $\mathrm{S}_{\circ}$ is the peak flux density, A and B are the lengths at which $\mathrm{V}(u, v)$ falls to 1/e of $\mathrm{S}_{\circ}$. We explicitly fit a Gaussian shape to our visibility functions because it is likely that the VLA samples the phase structure function below the "inner scale". Observations of several heavily-scattered objects show a quadratic dependence on $u$ and $v$ and also show a $\lambda^2$-dependence on the angular diameter (Wilkinson et al. 1988; Spangler & Gwinn 1990; Moran et al. 1990; Lo et al. 1993). These results are all consistent with the existence of an inner scale to the electron density fluctuations of $l_1 \sim 100 - 500$ km. The visibility coordinates $u$ and $v$ are modified to $u_{\phi}$ ($=u \sin\phi + v \cos\phi$) and $\mathrm{v}_{\phi}$ ($=u \cos\phi - v \sin\phi$) to account for the rotation of A and B by a position angle $\phi$, defined to be positive going North through East. $\mathrm{X}_u$ and $\mathrm{Y}_v$ are position offsets of the source from the pointing center. The real and imaginary components of the visibility were fitted separately to avoid the noise bias introduced when fitting the modulus of $\mathrm{V}(u, v)$. Two different non-linear least squares algorithms were employed as a check on the results.

Table 1 contains the results of each fit. Additional observational information pertaining to all sources in Table 1 can be found in the catalog of GC OH/IR stars by Lindqvist et al. (1991). The name of each source is given in column (1), followed by the relevant parameters from the fit and the estimated errors in columns (2), (3) and (4). The ratio $\theta_{maj}/\theta_{min}$ is the axial ratio $\eta$ and is given in column (4). The last column is the angular distance from SgrA*.

## 3. Results

Figure 1 illustrates the basic dataset used in the model fitting. The data have been averaged and normalized by their zero spacing flux density. Both OH 0.892+1.342 and OH 0.334−0.181 were observed on the same day and yet their visibility functions look very different. OH 0.892+1.342 is essentially a point source while OH 0.334−0.181 is significantly broadened. We attribute this decreasing amplitude as a function of interferometer baseline to interstellar scattering (see Paper I).

---

**Figure 1**

---

Inspection of Table 1 reveals several important features of this data. Both OH 0.892+1.342 and OH 359.762+0.120 were observed in Paper I, yielding diameters of 57±6 mas and 495±120 mas, respectively. Our fitted value for OH 359.762+0.120 ($\theta_{obs} \simeq \sqrt{\theta_{maj}\,\theta_{min}}$) is (marginally) consistent with Paper I but may in fact be slightly larger. This could simply reflect the lower signal-to-noise and more limited UV-coverage of the VLBA observations. OH 0.892+1.342 is essentially a point source at the range of interferometer baselines we sample with the VLA and so our fitting algorithm is unable to properly determine $\theta_{obs}$. We were unable to fit an elliptical Gaussian to the limited data for OH 0.319−0.040, so a circular fit is given instead. The new values of $\theta_{obs}$ in Table 1 are significantly in excess of the average scattering diameter of 525 mas reported in Paper I. The locations of the OH/IR stars and the properties of their scattering disks are shown in Figure 2.

---

**Figure 2**

---

None of these OH/IR stars is seen in projection against the highly-polarized radio filaments or arcs in the GC, identified as structures produced in a poloidal magnetic field (Yusef-Zadeh, Morris, & Chance 1984). In Paper I we determined that the site of enhanced scattering was approximately centered (angularly) on SgrA* with a radius of 15′. The outer radius was based on the fact that 7 sources within 15′ of SgrA* had large values of $\theta_{obs}$, whereas $\theta_{obs}$ dropped precipitously for OH0.190+0.036 and SgrB2, located 20′ and 45′ away from SgrA*, respectively. It appears that OH0.190+0.036, as speculated in Paper I, may indeed be a foreground object, as heavy scattering is seen in our new data for 4 OH/IR stars located about 25′ away from SgrA*. Two of these OH/IR stars are west of SgrA* and two are east of SgrA*. If we exclude OH0.190+0.036, there is no evidence for a gradient in the strength of scattering across the region. Thus the region of enhanced scattering still appears to be approximately centered on SgrA* where the outer radius is at least 25 arcmin and may in fact be as large as 45 arcmin.

Of potentially greater importance than the large values of $\theta_{obs}$, is that we measure non-circular scattering disks with axial ratios of order 3:1. These do not appear to be an artifact of the fitting process as the non-circularity is seen in the raw data. Apart from extragalactic sources which are seen in projection against the solar wind (Armstrong et al. 1990), we know of only one confirmed case of a non-circular scattering disk and that is SgrA*, for which Lo et al. (1993) measure an angular diameter at 3.6 cm of 17.5 mas $\times$ 8.6 mas, $\phi = 87 \pm 5^\circ$ and $\eta$=2.0. Similar results are obtained by Alberdi et al. (1993) at 1.35 cm, with $\theta_{obs}$=2.58 mas, $\phi = 79 \pm 6^\circ$ and $\eta$=2.0. The more modest anisotropy of the galactic source Cyg X-3 is $\eta \approx 1.3$ (L. Molnar, R. Mutel, M. Reid, & K. Johnston, unpublished; R. Narayan, private communication). In §4 we will explore what these new observations tell us about the location and properties of the enhanced turbulence towards the Galactic center.

## 4. Discussion

### 4.1. PROPERTIES OF THE SCATTERING MEDIUM

In Paper I we determined that $\bar{\theta}_{obs} \simeq 525$ mas for sources in a $15'$ region approximately centered on SgrA*. Our new results clearly show that the region of enhanced scattering is more extended about SgrA* than was previously known, with an angular radius of at least $25'$ centered on SgrA*. Large, non-circular scattering disks are also measured for the majority of sources. Some of the angular diameters that we have measured are 2-3 times larger than had been seen at the GC before. Both the strength of the scattering and the non-circular scattering disks (see §4.2) argue forcefully for a single region or "clump" of intense turbulent plasma. In Paper I we considered two possibilities for the location of this clump; either it is local to the GC and a consequence of the extreme properties of this region or else it is located along the line of sight, a random superposition and unrelated to the GC. While the first alternative is appealing, present arguments in its favor are largely phenomenological. A point in favor of the second alternative is that these random clumps are expected to be found throughout the disk of the Galaxy with a mean free path of approximately 8 kpc in order to gain agreement with the data for models of the large-scale distribution of turbulent ionized gas in the Galaxy (Cordes et al. 1991). An example of a scattering region not associated with the galactic center is the HII complex NGC6334, which scatters masers within the complex and scatters a background source by a large amount (Moran et al. 1990).

Either alternative can explain the new, larger angular diameters that we measure simply by allowing for variations of the strength of scattering. In Paper I $\theta_{obs}$ was seen to change along random lines of sight by a factor of 3-5 on angular scales of only $25'$ for sources believed to be at the same distance. We interpreted these variations as defining a physical size to the clumps of 50(L/8.5 kpc) pc, where L is the distance to the clump. If the scattering medium is located close to the GC the the spread in scatter sizes may also originate from the *range* in distances of the OH/IR stars behind the screen. As we know from dynamical considerations that the cluster of OH/IR stars at the GC has a $1/e$ depth of $\approx 50$ pc (Lindqvist et al. 1992). Given this stellar distribution and the observed distribution of scatter broadened images, it can be shown that a simple model of a homogeneous scattering screen near the GC can produce the observed scatter in angular sizes.

The GC model has one severe problem. It was shown in Paper I that the same density irregularities which produce angular broadening will contribute to the free-free absorption, by an amount that is strongly dependent on the location of the scattering material from the GC. Since SgrA* is not absorbed at centimeter wavelengths, either the outer scale is much smaller ($l_\circ << 1$ pc) or the temperature is much higher (T$>> 10^4$ K) than we assumed in Paper I.

Previous estimates for the outer scale have been based on comparisons of the scattering and emission measures toward high galactic latitudes, indicating an outer scale of $\sim 100$ pc (Cordes et al. 1993). Studies of Faraday rotation (Simonetti & Cordes 1986; Lazio et al. 1990; Clegg et al. 1992) also suggest outer scales that exceed 1 pc. The outer scale for heavily scattered lines of sight may not be so large, however. Frail, Kulkarni & Vasisht (1993) have shown that for other heavily scattered lines of sight in the Galaxy the data require that $l_\circ \leq$0.01 pc. In order for the turbulent gas to be located within 50 pc of the GC (for T=$10^4$ K) the outer scale would have to be only about 40 AU. Since there is a stronger dependence of the screen placement with gas temperature ($T^{-3/4}$ versus $l_\circ^{1/3}$), a hot scattering medium is suggested. The detection of a strong 6.7 keV line at the GC with the Ginga satellite (Yamauchi et al. 1990) has been interpreted by Spergel & Blitz (1992) as arising from a hot ($10^8$ K), diffuse (0.03-0.06 cm$^{-3}$) gaseous component within the inner 500 pc of the GC. Such gas would escape the Galaxy as a high velocity wind constrained only by the magnetic field. This flow could provide the required turbulence but it runs counter to theoretical expectations that density variations of the necessary magnitude cannot be supported in high temperature plasmas (Cesarsky 1980), and the limited evidence that the high temperature plasma surrounding the Sun is relatively quiescent (Phillips & Clegg 1992). Alternatively this hot wind might drive turbulence in the more dense gas at the GC without violating the emission measure constraint.

### 4.2. ANISOTROPIC SCATTERING

Non-circular disks are thought to be due to anisotropic scattering and they indicate that there exists a preferred direction for the density $\delta n_e$ irregularities (Cordes et al. 1984). This strongly suggests that the observed broadening comes from a small number of regions of enhanced turbulence since the superposition of many random turbulent cells would result in asymptotically circular scattering disks. Scattering from a single region with uniform magnetic field might yield even larger axial ratios, comparable to those $(5-15)$ seen in the solar wind within 5 solar radii of the Sun (Armstrong et al. 1990). The directionality could be provided by a velocity flow or a magnetic field, either one of which "stretches" out the density irregularity spectrum parallel to the velocity or field vector. In general, the direction of maximum elongation of the image will be *perpendicular* to this vector. Refractive effects can also cause elongation of a circular image (Romani, Narayan & Blandford 1986).

Higdon (1986) discusses the physics of magnetically elongated turbules such as might arise in some regions of the Galaxy. We note, also, that the parameters of the $10^8$ K gas discussed by Spergel & Blitz, combined with a magnetic field strength of 1 mG imply a ratio of magnetic to gas pressure in the range of $\sim 50-100$, consistent with the existence of elongated microstructure along the magnetic field. The thermal speed ($\sim 10^3$ km s$^{-1}$) corresponds to a proton gyro radius of $\sim 100\ B_{\rm mG}^{-1}$ km, comparable to the inner scale that appears relevant to strong scattering regions.

Interior to a 200 pc region about the GC there are several measurements that suggest that the magnetic field is large (B$>$ 1 mG) and that it is predominantly poloidal (i.e. perpendicular to the galactic plane) (Sofue 1989, Yuseh-Zadeh 1989). Unlike in most of the Galactic disk, the magnetic field at the GC has an energy density comparable to the gaseous components and therefore exerts a strong influence on the dynamics of ionized gas (Heyvaerts 1989). If the plasma turbulence were to arise from a region close to the GC then the non-circular scattering disks would trace the magnetic field direction if the field is uniform along the line of sight. On the other hand, it is possible that the scattering disks shown in Figure 2, which do not exhibit any clear association with inferred field directions, result from the cumulative effects from a range of field directions. Alternatively, the field may be nearly uniform over the scattering region relevant to each source, but twisting of the field makes the orientation of the scattering disks appear unstructured.

## 5. Conclusions

We have shown that circumstellar OH/IR maser sources in the galactic center are scattered by electron density variations that are evidently anisotropic. The level of scattering, as quantified by the angular diameter at 1.6 GHz, is approximately 1 arc sec and exceeds, by a factor of $\sim 2$, that of the GC source SgrA*. More importantly, the angular diameters at 1.6 GHz exceed those of many galactic sources by one to several orders of magnitude. For example, the pulsar 1933+16 shows moderate scattering of 15 mas at 0.33 GHz, corresponding to 0.5 mas at 1.6 GHz (Gwinn et al. 1993). Its distance (7.9 kpc) is nearly equal to that of the GC. The most heavily scattered pulsar (1849+00) behind the supernova remnant G33.6+0.1 shows pulse broadening that is consistent with angular broadening of 0.14 arc sec at 1.6 GHz.

Both the strength of the scattering and the non-circular scattering disks suggest that the scattering of sources in the direction of the GC is dominated by regions that are small compared to the total path length. The elongated scattering disks of the GC sources also suggest that, in this scattering region, the magnetic field is large enough to influence the shapes of irregularities and that it is sufficiently uniform in the scattering material to impart a preferred direction to the scattering.

The present data and also the results of Paper I do not determine uniquely the location, along the line of sight, of the material that scatters radiation from the GC sources. However, the data presented suggest that the scattering material is at least angularly centered on the GC. Combined with the possibility that the scattering material might correspond to (or be influenced by) the $10^8$ K gas detected by Ginga (Yamauchi et al. 1990), whose free-free absorption would not be large, the available evidence seems to favor the placement of the strong scattering region very near the GC.

As noted in Paper I, the presence of strong scattering material in the GC can be tested by making angular diameter observations of background extragalactic sources. Such sources, at distances much greater than the distance of the material from us (in contrast to the OH/IR sources and SgrA* in the GC), would show correspondingly larger scattering diameters ($\theta_{obs} \simeq 2'$). Further work should also include imaging of additional OH/IR stars to map out the scattering strength and degree of anisotropy. Infrared polarization measurements would also be useful for determining the orientation of grains by magnetic fields in the GC (Jones & Gehrz 1990; Jones et al. 1992).

**Acknowledgements**

The authors thank D. Backer and R. Narayan for useful discussions. DAF wishes to thank K. Desai for the use of his fitting program. JMC received support from NASA and the NSF, as well as from the National Astronomy and Ionosphere Center, which operates the Arecibo Observatory under a cooperative agreement with the NSF.

**Figure Captions**

**Figure 1.** The normalized real component of the visibility function for the OH/IR stars OH 0.892+1.342 and OH 0.334−0.181 as a function of interferometer baseline (expressed in thousands of wavelengths, where $\lambda$ = 18.61 cm). The data have been bin-averaged because the rms noise on a single 30 sec visibility measurement is about 75 mJy. OH 0.892+1.342 is point-like at the range of baselines sampled here.

**Figure 2.** The distribution of the OH/IR stars towards the Galactic center and the properties of their scattering disks. The ellipses represent the locations, sizes and orientations of the scattering disks as given in Table 1. The angular sizes have been scaled up for easier study. The OH/IR stars are superimposed on a 327 MHz radio image of the Galactic center by Pedlar et al. (1989). It is used here with permission of the authors.